\documentclass[pre,superscriptaddress,twocolumn,showpacs]{revtex4}
\usepackage{graphicx,bbm}
\usepackage{hyperref}
\usepackage{amsmath}
\usepackage{amsmath}
\def\etal#1{ {\em et al.}}
\def\tit#1{}

\begin{document}
\pagenumbering{arabic}

\title{Emerging spectra of singular correlation matrices under small power-map deformations}
\author{Vinayak}
\email{vinayaksps2003@gmail.com}
\affiliation  {Instituto de Ciencias F\' isicas, Universidad Nacional Aut\' onoma de M\' exico, C.P. 62210 Cuernavaca, M\' exico}

\author{Rudi Sch\"{a}fer}
\email{rudi.schaefer@uni-due.de}
\affiliation {Faculty of Physics, University of Duisburg-Essen, Germany}

\author{Thomas H. Seligman}
\email{seligman@ce.fis.unam.mx}
\affiliation  {Instituto de Ciencias F\' isicas, Universidad Nacional Aut\' onoma de M\' exico, C.P. 62210 Cuernavaca, M\' exico}
\affiliation {Centro Internacional de Ciencias, C.P. 62210 Cuernavaca, M\' exico}
\date{\today}

\begin{abstract}
Correlation matrices are a standard tool in  the analysis of the time evolution of complex systems in general and financial markets in particular. Yet most analysis assume stationarity of the underlying time series. This tends to be an assumption of varying and often dubious validity. The validity of the assumption improves as shorter time series are used. If many time series are used this implies an analysis of highly singular correlation matrices. We attack this problem by using the so called {\it power map} which was introduced to reduce noise. Its non-linearity breaks the degeneracy of the zero eigenvalues and we analyze the sensitivity of the so emerging spectra  to correlations. This sensitivity will be demonstrated for uncorrelated and correlated Wishart ensembles.
\end{abstract}

\maketitle

\section{Introduction}
Correlation matrices for time series are an old subject of research \cite{Wilks,Muirhead,markowitz59,pafka04,b_mcneil05,Bio1, Bio2,Medc, Thomas2012,mantegna97,b_voit01,cooley01,pelletier05,b_bouchaud00} and Wishart \cite{Wishart} has introduced white noise (stochastic) time series as a null hypothesis \cite{comment} to identify {\it actual} underlying cross correlations. More recently there have been advances incorporating correlations in random matrix ensembles \cite{SenM,MousSimon,Burda,vp2010,guhr1, guhr2, vin2013}. Applications \cite{Finance1, Finance2,Finance3,Finance4,diverseAT, gene, Sesmic, Seba:03, vmarko} have been varied with an emphasis on financial time series \cite{mantegna97,b_voit01,cooley01,pelletier05}, both because of their importance and their ready availability. Correlations are seen in stock-exchange data, resulting in large eigenvalues which represent the overall market movement and the industrial sectors \cite{b_bouchaud00}.

Applications usually imply stationarity of the time series after eliminating some well known trends. The problem we wish to address results from the fact that we may well have a much larger number of time series than the number of time steps over which the time series can reasonably be considered as stationary. This situation leads to correlation matrices that are highly singular. The purpose of the present paper is to study the properties of such matrices in presence of correlations.

Correlations measure the degree of linear dependence between the stochastic components of different time series.
Estimation errors arise due to the finite length of time series on which correlations are estimated. It is essential to suppress the corresponding noise in correlation matrices to reveal the actual correlations. Various techniques are available \cite{b_bouchaud00,other_tech1, other_tech2, GuhrKabler2003,GuhrShafer2010}. Among these we shall use a recent and efficient one, namely the power map \cite{GuhrKabler2003,GuhrShafer2010,Schmitt2013} both for noise reduction and for a purpose, quite different from the one for which it was designed and more central to our paper. 

In the problem we wish to address we do not have long time series usual for the application of the power map. Yet we wish to use this map because of its non-linearity. The power map operates directly on the correlation matrix by the simple means of elevating the absolute value of the matrix elements to a power greater than one while conserving their phase. This will lift the degeneracy of the zero eigenvalues of the singular correlation matrix. The thus emerging spectrum gives us a handle to get more information from the eigenvalues without looking at the entire correlation matrix as in \cite{Thomas2012}. This may be useful because the correlation matrix contains considerable redundancies, if the number of time series is much larger than twice the length of the time series. Yet the first step will be to check, whether in such a context the power map will still provide the desired noise reduction (see \cite{GuhrShafer2010} as well); thus in the next section we shall test, in a simple model, whether noise reduction by means of the power map is effective for singular correlation matrices. Next we proceed to the main part of our results, which will be to analyze the behavior of Wishart ensembles (WE) of singular matrices \cite{MaslovZhang2001,janiknowak2003}, including correlated Wishart ensembles (CWE), under the power map deformations.

The analysis will be partially numerical and partially by linear response studies to obtain approximate analytical expressions for powers near identity. Maybe the most important finding is that the part of the spectrum we get by lifting the degeneracy at zero is sensitive to correlations and for powers near identity is well separated from the bulk, which in turn is only little affected. The main focus of this work will be centered on this part of the spectrum, which we call the {\it emerging spectrum}. 

In the next section we will give an example of the effectiveness of the power map for noise reduction in singular correlation matrices. The third section gives the framework in which we shall discuss correlated random matrix ensembles. In the following section we shall discuss the emerging spectra for a Wishart orthogonal ensemble without correlations comparing linear response analytics with numerics. In Sec. V we perform a similar analysis for two types of correlated Wishart ensembles though the analytic part is more restricted in this section. In Sec. VI we summarize our results and give an outlook on possible applications.

\section{The Power Map for Singular Correlation Matrices}
\begin{figure}
        \centering
               \includegraphics [width=0.47\textwidth]{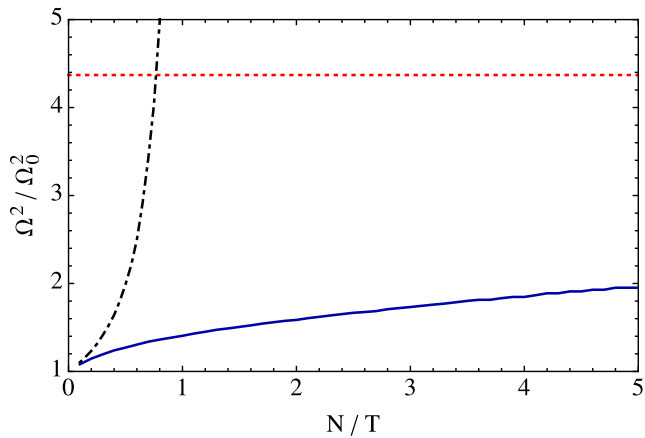}
                \caption{Portfolio variance $\Omega^2$ normalized to the minimal portfolio variance $\Omega_0^2$. The length $T$ of the time series was varied, the number of companies was fixed at $N=100$. Results are shown for sample correlations without noise reduction (black dashed-dotted line) and
for power-mapped correlations (blue solid line).The red dashed line at $\Omega^2/\Omega_0^2=4.37$ corresponds to a homogeneous portfolio.}
\label{pmbeyond}
\end{figure} 
{As the power map has mainly been discussed for regular matrices we start giving a brief discussion of previously unpublished ( see also \cite{GuhrShafer2010}) results for this case from the realm of econophyscs. In this section we shall use standard notations of econophysics as summarized in Appendix \ref{app:opt}.} Following Markowitz~\cite{markowitz59}, we consider a portfolio of $N$ stocks and wish to calculate the portfolio weights $w_\mathrm{opt}$ which minimize the portfolio variance
\begin{equation}
\Omega^2=w_\mathrm{opt}^t \Sigma w_\mathrm{opt} \ .
\end{equation}
In a model setting, we can calculate the minimal variance portfolio $\Omega_0^2$ using the model covariance matrix $\Sigma_0$. In practice, however, the covariance matrix has to be estimated using historical data of finite length $T$. The shorter the length $T$ of the time series, the noisier is the covariance estimation. Using noisy covariance matrices for portfolio optimization leads to very bad results, see Fig.~\ref{pmbeyond}. In this case, the portfolio variance $\Omega^2$ increases as $(1-N/T)^{-1}$, in accordance with literature~\cite{pafka03}. Clearly, it is necessary to improve the estimation of the covariance matrix to obtain better results. The variances of the single stocks can be estimated rather well on short time horizons due to a slowly decaying autocorrelation. The noise in the correlation coefficients $C_{kl}$ can be reduced effectively using the power map
\begin{equation}\label{pmap}
C^{(q)}_{kl}={\rm sign}\,C_{kl} \, \big| C_{kl} \big|^q \ .
\end{equation}
The results are presented in Fig.~\ref{pmbeyond}. The power map yields portfolio variances which are well below the homogeneous portfolio with all weights equal to $1/N$, even for $N>T$ where the correlation matrix becomes singular. The values for $q$ used in this study range from 1.1 to 2.4. For details of the simulation we refer to Appendix \ref{app:opt}.

\section{Random Correlation Matrices: Wishart and Correlated Wishart Ensembles}
In random matrix theory one studies basically three invariant Wishart ensembles \cite{Mehta}. In this paper we consider only Wishart orthogonal ensemble (WOE). Similarly we consider the correlated Wishart orthogonal ensemble (CWOE) among the three invariant CWEs. A Wishart matrix is defined as $\mathsf{C}=\mathsf{AA}^{t}/T$ where $\mathsf{A}$ is an $N\times T$ matrix and $\mathsf{A}^{t}$ is the transpose of $\mathsf{A}$. For WOE the matrix elements of $\mathsf{A}$ are real independent Gaussian variable with mean zero and a fixed variance $\sigma^{2}$. In the context of time series $\mathsf{C}$ may be interpreted as the correlation matrix, calculated over stochastic time series of time horizon $T$ for $N$ statistically independent variables. By construction $\mathsf{C}$ is a real symmetric positive semi-definite matrix. 

Using the Gaussian probability measure, the joint probability density (JPD) of the matrix elements of $\mathsf{A}$ can be written as
\begin{equation}\label{jpdajk}
P(\mathsf{A})\,d\mathsf{A}\propto \exp\left[-\text{Tr}\dfrac{\mathsf{AA}^{t}}{2\sigma^{2}}\right]d\mathsf{A},
\end{equation}
where $d\mathsf{A}$ is the infinitesimal volume in $N\times T$ matrix element space. For $T\ge N$, the JPD of the matrix elements of $\mathsf{C}$ is defined in $N(N+1)/2$-dimensional matrix element space \cite{Wishart,Wilks}, as
\begin{eqnarray}\label{jpdcjk}
P(\mathsf{C})\,d\mathsf{C}\propto (\text{det}\mathsf{C})^{[N(\kappa-1)-1]/2}\,\exp\left[-\dfrac{T}{2\sigma^{2}}\,\text{Tr}\,\mathsf{C}\right]\,d\mathsf{C},
\end{eqnarray}
where $\kappa=T/N$ and $d\mathsf{C}=\prod_{j> k}^{N}Sc_{jk}\prod_{j=1}^{N}dC_{jj}$ is the infinitesimal volume in matrix element space. The JPD of the eigenvalues of $\mathsf{C}$, $\lambda_{j}$, for $j=1,...,N$, may be obtained by transforming the variables to eigenvalue-eigenvector space and then integrating over the eigenvectors. It leads to 
\begin{eqnarray}\label{jpdigvl}
P(\lambda_{1},...,\lambda_{N})\propto \prod_{j=1}^{N}w(\lambda_{j})\prod_{j>k}^{N}|\lambda_{j}-\lambda_{k}|,
\end{eqnarray}
where $w(\lambda)=\lambda^{[N(\kappa-1)-1]/2}\exp[-N\kappa \lambda/2\sigma^{2}]$ is the weight function of the associated Laguerre polynomials {hence WOE is often referred to as Laguerre orthogonal ensemble in literature \cite{Mehta,Brody81,TGW:98,BenRMP:97}}. For $T< N$, $\mathsf{C}$ is singular and has exactly $(N-T)$ zero eigenvalues. The JPD of the matrix elements and the JPD of the eigenvalues in this case have been derived in \cite{MaslovZhang2001,janiknowak2003}, where again Laguerre weight functions describe JPD of the latter. The eigenvalue statistics of WOE are known in great detail \cite{Mehta}. {For instance,} the global statistics, i.e. the eigenvalue density, as well as the local statistics, i.e. the $n$-point {spectral} correlation functions for $n\geq2$, are known in terms of Laguerre polynomials. However, for $N,T\rightarrow\infty$ such that $\kappa\ge1$ and finite, the ensemble averaged eigenvalue density converges to the Mar\'{c}enko Pastur law \cite{marchenko}: 
\begin{equation}
\label{denmp}
\overline{\rho}(\lambda)=\kappa\dfrac{\sqrt{(\lambda_{+}-\lambda)(\lambda-\lambda_{-})}}{2\pi \sigma^{2}\lambda},
\end{equation}
where $\lambda_{\pm}=\sigma^{2}(\kappa^{-1/2}\pm 1)^2$ are the end points of the density. We use here and further down bar to denote ensemble averaging. Note that for the singular case, i.e. $\kappa<1$,  $\overline{\rho}(\lambda)$ in the above equation is normalized to $\kappa$ and not to $1$. Therefore, taking into account of $(N-T)$ zeros, for $\kappa\le1$ we write
\begin{equation}
\label{denmpkle1}
\overline{\rho}(\lambda)=\kappa\dfrac{\sqrt{(\lambda_{+}-\lambda)(\lambda-\lambda_{-})}}{2\pi \sigma^{2}\lambda}+(1-\kappa)\delta(\lambda).
\end{equation}
{When expressed in terms of unit average spacing the $n$-point correlation functions, for large matrices, }are the same as those for Gaussian orthogonal ensemble (GOE) \cite{Mehta}, and therefore WOE spectra have universal spectral fluctuations \cite{PandeyGhosh2001}. 

In the case of actual correlations among the matrix elements $A_{jk}$, one defines CWOE \cite{Wilks}. For instance, in order to take account of the correlation in the rows of $\mathsf{A}$ matrix one defines $\mathsf{C}=\xi^{1/2}\mathsf{BB}^{t}\xi^{1/2}$ where $\xi$ is a fixed positive definite matrix which takes account of nonrandom correlations and where the matrix elements $B_{jk}$ are independent Gaussian variables with zero mean and a fixed variance, just as the $A_{jk}$ defined for the WOE. We remark that in this definition $\overline{\mathsf{C}}=\sigma^{2}\xi$ while $\overline{\mathsf{A^{\text{t}}A}}/N$ is an identity matrix of dimension $T\times T$. Analytical treatment to CWOE \cite{SenM,Burda,vp2010, guhr1, guhr2} is much more difficult and involved than to WOE. However, some results are known in this case as well. For instance, analogous to the Mar\'{c}enko Pastur result, the resolvent or Stieltjes transform of the density is known as
\begin{equation}
\label{resCWE}
\overline{G}(z)=\left\langle
\dfrac{1}{z-\dfrac{\sigma^{2}}{\kappa}\left(\kappa-1+z\overline{G}(z)\right)\xi}
\right\rangle,
\end{equation}
where angular brackets denote the spectral averaging. The density $\overline{\rho}(\lambda)$ can be determined uniquely via inverse transformation:
\begin{equation}
\label{denCWE}
\overline{\rho}(\lambda)=\mp \dfrac{1}{\pi}\,\Im\,\overline{G}(\lambda\pm\text{i}\epsilon),
\end{equation}
where $\epsilon>0$ is infinitesimal. For finite $N$ and $T$, the density has been recently obtained \cite{guhr1, guhr2} by using the supersymmetric method. For higher order spectral correlations, only asymptotic result for the two-point function is known \cite{vp2010}. Spectral fluctuations of CWOE are not always universal; rather there are deviations in some cases, as noted for the number variance \cite{Brody81}, $\Sigma^{2}(r)$, for large spectral correlation length $r$. In some cases, however, a few eigenvalues are found to be separated from the bulk density. These eigenvalues often show collective behaviour, therefore, referred to as the collective modes. We omit further details here, however interested readers may find a detailed discussion on the spectral properties of CWEs in \cite{vp2010}.

\section{Estimation of the moments of emerging spectra for WOE in a linear response regime}

{The emerging spectra, we wish to study for the WOE, may be observed even for $q$ very near to identity. Before developing an analytic approach we mention a few important spectral properties of $\mathsf{C}^{(q)}$ as observed in simple numerical simulations. First we note that $\mathsf{C}^{(q)}$ is always real symmetric; thus it has real eigenvalues. However, for $q\ne1 $, it may have negative eigenvalues specially when $T$ is much smaller than $N$. The density function of the eigenvalues of $\mathsf{C}^{(q)}$, appears on two well separated supports. The first one is close to zero while the other is close to the support defined by the Mar\'{c}enko-Pastur density (\ref{denmp}). The former results from the breaking of degeneracy of the zero eigenvalues of $\mathsf{C}$ while the latter is due to small corrections to non-zero spectrum. As we increase the power to values usually used for noise reduction, the two supports begin to overlap.

We illustrate some of these remarks with numerics of the WOE case. Consider a $1024\times 512$ random matrix $\mathsf{A}$ where the matrix elements are independent Gaussian variables with mean zero and variance one. Let the exponent of the power map (\ref{pmap}) be close to one, say, $q=1.001$. In Fig. \ref{dencq}(a) we show the density of the emerging spectra and in Fig. \ref{dencq}(b) we show the density of the eigenvalues near the Mar\'{c}enko-Pastur density and is actually very close to the latter. 

\begin{figure}
        \centering
               \includegraphics [width=0.5\textwidth]{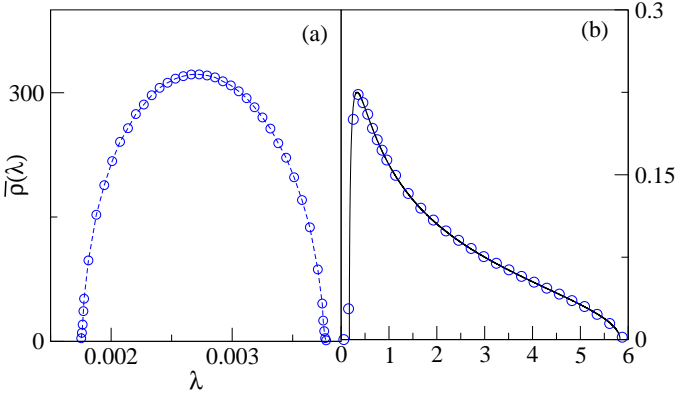}
                \caption{Density of eigenvalues of $\mathsf{C}^{(q)}$ where $q=1.001$, $N=1024$ and $\kappa=1/2$ for the WOE case. In Fig. \ref{dencq}(a) we show the density of emerging spectra while in Fig. \ref{dencq}(b) we show the density of the non-zero eigenvalues which is closely described by the Mar\'{c}enko Pastur law (\ref{denmp}) shown by a solid line. Both densities are shown on different scales. The density in Fig. \ref{dencq}(a) is normalized to $1-\kappa$ while the density in Fig. \ref{dencq}(b) is normalized to $\kappa$. Note that the density in Fig. \ref{dencq}(a) is not quite symmetric.}
\label{dencq}
\end{figure}

Analytically, we derive an estimate of the first two moments of the density shown in Fig. \ref{dencq}(a). As we wish to make an expansion around $q=1$ we introduce the small parameter  \ $\alpha=(q-1)$ and define $\mathsf{C}^{(\alpha)}\equiv\mathsf{C}^{(q)}$ as defined in Eq. (\ref{pmap}). For small $\alpha$, $\mathsf{C}^{(\alpha)}$ may be expanded as }
\begin{eqnarray}\label{LRC}
C_{jk}^{(\alpha)}&=&C^{(0)}_{jk}\exp\left[\dfrac{\alpha}{2}\text{ln}[(C^{(0)}_{jk})^{2}\right]
\nonumber\\
&=& C_{jk}+\dfrac{\alpha}{2}\,C_{jk}\,\text{ln}(C_{jk}^{2})\left[1+\mathcal{O}\left(\alpha\right)\right],
\end{eqnarray}
where in the second equality we drop the superscript, using from now on $\mathsf{C}$ for $\mathsf{C}^{(0)}$. Next we expand the eigenvalues $\lambda_{j}(\alpha)$, of $\mathsf{C}^{(\alpha)}$, as
\begin{equation}\label{asigvl}
\lambda_{j}(\alpha)=\lambda_{j}(0)+\alpha\,(\delta\lambda_{j})[1+\mathcal{O}(\alpha)].
\end{equation}
Here the $\lambda_{j}(0)'$s are the eigenvalues of $\mathsf{C}$, for $j=1,..., N$ and the $\alpha(\delta\lambda_{j})'$s are the leading order corrections coming from the power mapping. For a short time horizon, i.e. $T<N$, $\lambda_{j}(0)=0$ for $j\le N-T$. For small $\alpha$, we assume that the statistics of the relative changes in the eigenvalues is dominated by the linear term. Bearing this in mind, we derive estimates for the moments of the $\alpha(\delta\lambda_{j})'$s in the linear response regime. We refer to $\alpha(\delta\lambda_{j})$ as the eigenvalues of the emerging spectrum, for $j\le N-T$, otherwise as corrections of the  non-zero eigenvalues or non-zero eigenvalue corrections. We consider $\alpha> 0$. 

 Since we are mainly interested in spectral statistics, specially in the first few spectral moments, it is sufficient to work in the matrix element space using linear response theory as explained above. Perturbation theory for all individual eigenvalues would yield all moments and therefore the density\cite{Wegmann}, but the effort seems prohibitive.

We now define the moments, for integer $n\ge0$, as
\begin{eqnarray}\label{moms}
\overline{m}_{n}&=&\dfrac{1}{N}\overline{\text{Tr}\,\left[\mathsf{C}^{(\alpha)}\right]^{n}}=\dfrac{1}{N}\overline{\sum_{j=1}^{N}(\lambda_{j}(\alpha))^{n}},
\nonumber\\
\label{mommp}
\overline{m^{\text{mp}}}_{n}&=&\dfrac{1}{N}\overline{\text{Tr}\,\left[\mathsf{C}\right]^{n}}=\dfrac{1}{N}\overline{\sum_{j=1}^{N}(\lambda_{j}(0))^{n}},
\end{eqnarray}
where the superscript $\{\text{mp}\}$ denotes the moments which in the large $N$ limit will tend to the moments of the Mar\'{c}enko Pastur density (\ref{denmp}). Next, for {all} the  eigenvalue corrections, we define the moments as
\begin{eqnarray}\label{delm}
\overline{\delta m_{n}}&\equiv&\dfrac{\alpha^{n}}{N}\overline{\sum_{j=1}^{N}\,(\delta\lambda_{j})^{n}}.
\end{eqnarray}
Note that in the linear response regime $\overline{\delta m_{n}}$ may also be estimated by 
\begin{eqnarray}\label{avCq}
\overline{\delta m_{n}}
&\simeq&\text{$\alpha^{n}$ term in}
\dfrac{1}{N}\left[\overline{\text{Tr}\,\left(\mathsf{C}^{(\alpha)}\right)^{n}}-\overline{\text{Tr}\,\left(\mathsf{C}\right)^{n}}\right].
\end{eqnarray}
Defining $\overline{\delta m^{(0)}_{n}}$ as the moments of emerging spectra and $\overline{\delta m^{\text{mp}}_{n}}$ as the moments of non-zero eigenvalue corrections, we write
\begin{eqnarray}\label{m0pm1p}
\overline{\delta m}_{n}=
\overline{\delta m^{(0)}_{n}}+\overline{\delta m^{\text{mp}}_{n}}.
\end{eqnarray}
{An important, though almost trivial remark is that $\overline{\delta m^{\text{mp}}_{n}}\equiv\overline{\delta m_{n}}$ for $T\ge N$.}

To calculate the averages in (\ref{avCq}) one needs the all the moments of $C_{jk}$. Using the JPD of $A_{jk}$, given in (\ref{jpdajk}), these can be derived. We find
\begin{eqnarray}\label{momcjke}
\overline{(C_{jk})^{2n}}
&=&
\delta_{jk}\left(\dfrac{2\sigma^{2}}{T}\right)^{2n}\dfrac{\Gamma(2n+T/2)}{\Gamma(T/2)}+
\left(1-\delta_{jk}\right)
\nonumber\\
&\times&
\left(\dfrac{\sigma^{2}}{T}\right)^{2n}\,
\dfrac{\Gamma(2n+1)}{\Gamma(n+1)}
\dfrac{\Gamma(T/2+n)}
{\Gamma(T/2)},\\
\nonumber\\
\label{momcjko}
\overline{(C_{jk})^{2n+1}}
&=&
\delta_{jk}\left(\dfrac{2\sigma^{2}}{T}\right)^{2n+1}\,\dfrac{\Gamma(2n+1+T/2)}{\Gamma(T/2)},
\end{eqnarray}
for integer $n\ge0$ {and with arbitrary $\sigma^{2}$ (which of course is set to be equal one for correlation matrices; $\sigma^{2}=1$).

Using Eq. (\ref{LRC}) in Eq. (\ref{avCq}) for $n=1$, we estimate the first moment of the corrections, as
\begin{eqnarray}\label{delm1}
\overline{\delta m_{1}}&\simeq&\dfrac{\alpha}{2N}\sum_{j=1}^{N}\overline{C_{jj}\text{log}[(C_{jj})^{2}]}
\nonumber\\
&\simeq&\dfrac{\alpha}{2N}\sum_{j=1}^{N}\dfrac{d}{dh} \overline{(C_{jj})^{2h+1}}\Big|_{h=0}.     
\end{eqnarray}
In the above equation we first use Eq. (\ref{momcjko}) for $\overline{(C_{jj})^{2h+1}}$. Next, we take the derivative with respect to $h$ and finally set $h=0$. We then obtain}
\begin{equation}\label{delm2}
\overline{\delta m_{1}}=\alpha\left[\text{log}\left(\dfrac{2}{T}\right)+\Psi(1+T/2)\right],
\end{equation}
where $\Psi(x)$ is the digamma function. This equality is valid to leading order in $\alpha$. Similarly for the second moment we obtain
\begin{eqnarray}
\overline{\delta m_{2}}&=&\alpha^{2}
\left(1+\dfrac{2}{T}\right)
\Bigg\{
\left(\text{log}\dfrac{2}{T}+\Psi(2+T/2)\right)^{2}
\nonumber\\
&+&\Psi'(2+T/2)
\Bigg\}+
\dfrac{\alpha^{2}}{\kappa}\left(1-\dfrac{1}{N}\right)
\Bigg\{
\bigg(-\text{log}(T)
\nonumber\\
&+&
1-\dfrac{\gamma}{2}+\dfrac{\Psi(1+T/2)}{2}\bigg)^{2}
+\dfrac{\Psi^{'}(1+T/2)}{4}
\nonumber\\
&-&
1+\dfrac{\pi^{2}}{8}
\Bigg\},
\end{eqnarray}
where $\Psi'(x)=d\Psi(x)/dx$. The large $T$ asymptotic of $\overline{(\delta m_{1})}$ and $\overline{(\delta m_{2})}$ may be evaluated as
\begin{eqnarray}\label{m1m2ap}
\overline{\delta m_{1}}&\sim&\dfrac{\alpha}{T},
\nonumber\\
\overline{\delta m_{2}}&\sim&\dfrac{\alpha^{2}}{4\kappa}\left([\text{log}(T)+c_{1}]^{2}+c_{2}\right),
\end{eqnarray}
where $c_{1}=\gamma+\text{log}(2)-2= -0.729637...$ and $c_{2}=\pi^{2}/2-4=0.934802...$, for $\gamma$ being the Euler constant. Note that the density defined by $\overline{\delta m}_{n}$ is doubly peaked for $T<N${; one peak corresponds to the eigenvalues of emerging spectra and the other to the non-zero eigenvalue corrections. The former vanishes as $T\rightarrow N$.}

{We note that the moments $\overline{m^{\text{mp}}}_{n}$ depend on $\kappa$ only; see Eq. (\ref{denmp}).} Furthermore the moments defined in (\ref{m0pm1p}) might also have $T$-dependency, as we show below. Also estimating either of the moments $\overline{\delta m_{p}^{(0)}}$ or $\overline{\delta m_{p}^{(\text{mp})}}$ is in no way straightforward. However, in the linear response regime we may still assume that the ensemble averaged density of the non-zero eigenvalue corrections could be approximated by a rescaled Mar\'{c}enko Pastur density. Moreover, we allow {this scale to assume negative values as well, as there is no reason for the non-zero eigenvalue corrections  to be strictly positive}. We note that this density should be normalized to $1$ for $\kappa\ge1$ and to $\kappa$ otherwise. {We also use a shifting-parameter which completes the linear equation. We finally make an ansatz for this density,}
\begin{equation}\label{denansatz}
\overline{\rho}_{1}(\delta\lambda)=\kappa\dfrac{\sqrt{(\delta\lambda_{+}-\delta\lambda)(\delta\lambda-\delta\lambda_{-})}}{2\pi (\delta\lambda-r)s}.
\end{equation} 
Here $\delta\lambda_{\mp}=s(\kappa^{-1/2}\pm 1)+r$ while $r$ and $s$ are the shifting and scaling parameters respectively. We fix these parameters by calculating the first two moments of the density. For the first moment we obtain
\begin{equation}\label{m1st}
\overline{\delta m^{\text{mp}}_{1}}=\begin{cases}
s+r, & \text{for $\kappa\ge1$},
\\
s+\kappa \,r, & \text{for $\kappa\le1$}.
\end{cases}
\end{equation}
Similarly the second moment is given by
\begin{equation}\label{m2st}
\overline{\delta m^{\text{mp}}_{2}}=
\begin{cases}
\left(1+\dfrac{1}{\kappa}\right)s^{2} -r^{2}+2r\overline{(\delta m_{1})},& \text{for $\kappa\ge1$},
\\
\left(1+\dfrac{1}{\kappa}\right)s^{2} -r^{2}\kappa+2r\overline{(\delta m^{\text{mp}}_{1})}, & \text{for $\kappa\le1$}.
\end{cases}
\end{equation}
Inverting the above relations, we write
\begin{equation}\label{sm1m2}
s=\begin{cases}
-\sqrt{\left[\overline{\delta m_{2}}-\overline{\delta m_{1}}^{2}\right]\kappa},& \text{for $\kappa\ge1$},
\\
-\sqrt{\overline{\delta m^{\text{mp}}_{2}}-\dfrac{\overline{\delta m^{\text{mp}}_{1}}^{2}}{\kappa}},& \text{for $\kappa\le1$},
\end{cases}
\end{equation}
and
\begin{equation}\label{tm1m2}
r=\begin{cases}
\overline{\delta m_{1}}-s,& \text{for $\kappa\ge1$},
\\
\dfrac{\overline{\delta m^{\text{mp}}_{1}}-s}{\kappa},& \text{for $\kappa\le1$}.
\end{cases}
\end{equation}
For $\kappa\ge1$ the two parameters are given by $\overline{\delta m_{1}}$ and $\overline{\delta m_{2}}$ {while these moments behave smoothly as a function of $\kappa$, even at $\kappa=1$. On the other hand, we expect a non-smooth behaviour in $\overline{\delta m^{\text{mp}}_{1}}$ and $\overline{\delta m^{\text{mp}}_{2}}$ at $\kappa=1$, as all the moments of the emerging spectra become zero for $\kappa\ge1$}. We now use results (\ref{m1m2ap}) and (\ref{sm1m2}, \ref{tm1m2}) to write asymptotic forms of $s$ and $r$ for $\kappa\ge1$. For large $T$, we obtain $N$-independent behaviour for both the quantities, e.g.,
\begin{eqnarray}\label{stap}
s&\sim& -\dfrac{\alpha}{2}\sqrt{[\text{log}(T)+c_{1}]^{2}+c_{2}},
\nonumber\\
r&\sim&
\alpha\left(\dfrac{1}{T}+\dfrac{\sqrt{[\text{log}(T)+c_{1}]^{2}+c_{2}}}{2}\right).
\end{eqnarray}
Exploiting the $N$-independence in these results, we may extrapolate them for $\kappa\le 1$. {Note that $s$ should be negative for $\alpha>0$ while it should be a positive quantity for $\alpha<0$}. Now, the estimation of $\overline{\delta m^{\text{mp}}_{1}}$ and $\overline{\delta m^{\text{mp}}_{1}}$ is straightforward, e.g.,
\begin{eqnarray}\label{m11m12ap}
\overline{\delta m^{\text{mp}}_{1}}&=&\kappa\overline{\delta m_{1}}+s(1-\kappa),
\nonumber\\
\label{delm22}
\overline{\delta m^{\text{mp}}_{2}}&=&\kappa\overline{\delta m_{2}}-\kappa\overline{\delta m_{1}}^{2}+\dfrac{\overline{(\delta m^{\text{mp}}_{1})}^{2}}{\kappa}.
\end{eqnarray}
Using (\ref{m0pm1p}), estimation of the moments of the emerging spectra becomes trivial. For large $T$ and $\kappa\le1$, we get
\begin{eqnarray}\label{m01m02ap}
\overline{\delta m^{(0)}_{1}}&=&-s(1-\kappa),
\nonumber\\
\overline{\delta m^{(0)}_{2}}&=&s^{2}(1-\kappa).
\end{eqnarray} 
We must mention that for small values of $\kappa$, the error in our approach becomes large and linear response theory fails.  

\begin{figure}
        \centering
               \includegraphics [width=0.5\textwidth]{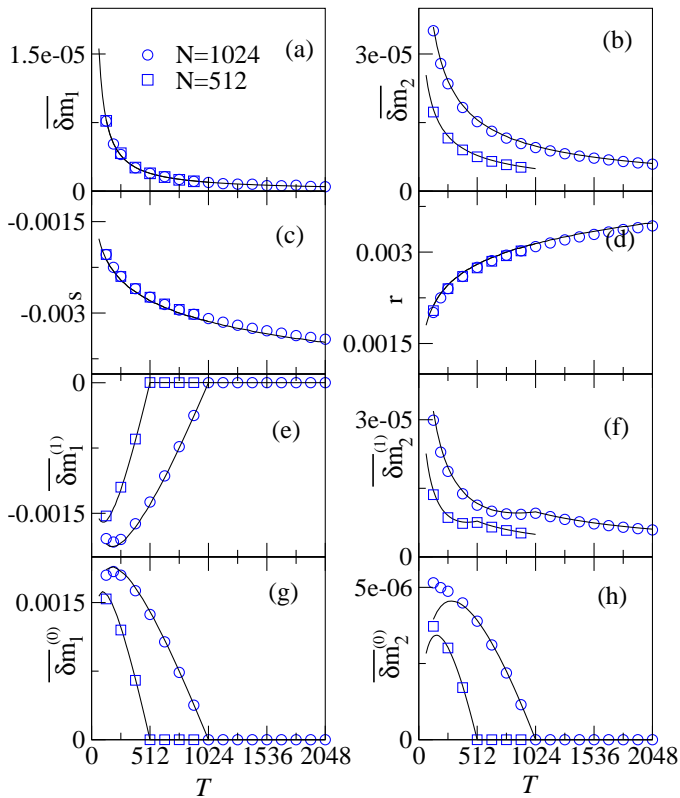}
                \caption{{Comparison of theory with simulations for WOE where $T$ is the variable. Solid lines are used to represent theoretical results and symbols are used to represent numerics.} In Figs. \ref{m1m2}(a) and \ref{m1m2}(b) we compare first two moments $\overline{\delta m_{1}}$ and $\overline{\delta m_{2}}$ obtained from the numerical simulations with the theory (\ref{m1m2ap}). In Figs.  \ref{m1m2}(c) and  \ref{m1m2}(d) we compare numerical results of scaling and shifting parameters $s$ and $r$, obtained from Eqs. (\ref{sm1m2}, \ref{tm1m2}) with theory (\ref{stap}). In Figs. \ref{m1m2}(e) and \ref{m1m2}(f) we compare numerical results of $\overline{\delta m^{\text{mp}}_{1}}$ and $\overline{\delta m^{\text{mp}}_{2}}$ with theory (\ref{m11m12ap}). Finally, in Figs. \ref{m1m2}(g) and \ref{m1m2}(h) we compare $\overline{\delta m^{(0)}_{1}}$ and $\overline{\delta m^{(0)}_{2}}$ with the theory (\ref{m01m02ap}).}
\label{m1m2}
\end{figure}

In Fig. \ref{m1m2} we have compared our theoretical estimates with numerical results for fixed $N=512$ and $N=1024$.  For the numerical results we have used $\overline{\delta m_{n}}=\overline{[\sum_{j=1}^{N}(\Delta\lambda_{j})^{n}]/N}$ where $\Delta\lambda_{j}=\lambda^{(\alpha)}_{j}-\lambda_{j}$; the $\Delta\lambda_{j}$'s are the eigenvalues of the emerging spectra for $j\le N-T$ otherwise these are the non-zero eigenvalue corrections. Similarly, for $\overline{\delta m^{(0)}_{n}}$ and $\overline{\delta m^{\text{mp}}_{n}}$ we use $\Delta\lambda_{j}$ summed respectively for $0<j\le N-T$ and $N-T<j\le N$. As shown in the graph that our theory gives reasonable account for the numerical results. The non-smooth behaviour in the moments of non-zero eigenvalue corrections or of the emerging spectra are well described by our theory. {As we see in Figs. \ref{m1m2}(c) and \ref{m1m2}(d) scaling and shifting parameters are indeed independent of $N$. For the first moment we see a couple of interesting features in Figs \ref{m1m2}(e) and \ref{m1m2}(g), e.g.  ($i$) the two curves are almost symmetric about zero, ($ii$) $\overline{\delta m^{\text{mp}}_{1}}$ reaches a minimum as $\overline{\delta m^{(0)}_{1}}$ reaches a maximum. Our theory for small $T$, however, fails for the second moment for emerging spectra. This is the regime where our linear response theory fails for higher order moments.}

\section{Emerging Spectra of CWOE}

Correlated ensembles are more difficult to handle but important results have been obtained for the CWOE case \cite{vp2010}. We use such ensembles to see how large the deviations from our null-hypothesis are and in particular if the power map is sensitive to the presence of correlations. We will have to rely mainly on numerical results, not only because the combined inherent difficulty of correlated ensembles and the power map, make analytical progress very hard but also because the number of ways we can choose the true correlations. 

In a first example we consider, $\xi$ is block diagonal. It has $\ell$ blocks of dimension $N_{i}, i=1...\ell$ such that $\sum_{i}^{\ell}N_{i}=N$. For the $i$'th block we have $\xi_{jk}=\delta_{jk}+(1-\delta_{jk})\,c_{i}$. Matrix elements of the off-diagonal blocks are $0$. Such examples qualitatively correspond to the 'factor model' in quantitative finance \cite{Noh2000,GuhrKabler2003,GuhrShafer2010}, where the off-diagonal blocks have small entries. In a second example, $\xi$ has smooth band structure, e.g., $\xi_{jk}=c^{|j-k|}$. {This model may have applications in time series analysis of any system, where we suspect short range correlations to dominate, while long range correlations are suppressed, or where short range correlations are unavoidable, while long range correlations are of interest; in the latter case we have an improved null hypothesis!} Without loss of generality in both cases we consider $c\ge0$. Naturally the set of interesting correlations is much bigger and may be very dependent on the problem. One example would be power law decay with its possible implications for critical statistics \cite{LombardiTHS, verbaarschot} will be investigated later. Yet we will leave this and other interesting examples for future study.

\subsection{Ensemble correlation matrix with block-diagonal structures}

The simplest example for which we can readily obtain analytical results is $\ell=1$. In this example $\xi$ is a dense matrix and its eigenvalues $\xi_{j}$ are simply given by $\xi_{j}=(1-c)$ for $j=1,...,N-1$ and $\xi_{N}=Nc+1-c$. For this spectrum, Eqs. (\ref{resCWE}, \ref{denCWE}) yields the density \cite{vp2010}
\begin{eqnarray}
\label{eqdeneqc}
\overline{\rho}(\lambda)&=&\overline{\rho'}(\lambda)+\delta\left(\lambda-\dfrac{(Nc+1-c)(Nc\kappa+1-c)}{Nc\kappa}\right),
\nonumber\\
\overline{\rho'}(\lambda)&=&
=\kappa\dfrac{\sqrt{(\lambda_{+}-\lambda)(\lambda-\lambda_{-})}}{2\pi (1-c)\lambda},
\end{eqnarray}
where $\lambda_{\pm}=(1-c)(\kappa^{-1/2}\pm 1)^2$. {Similarity of $\overline{\rho'}(\lambda)$ with the Mar\'{c}enko-Pastur law (\ref{denmp}) is evident here, with the only difference of a factor of $(1-c)$ in the place of $\sigma^{2}$ in (\ref{denmp})}. The delta function appears in the above result as long as $c\ge (N\sqrt{\kappa})^{-1}$. It is known from the work of Baik {\it et al} \cite{Baik:05} that the density of such separated individual eigenvalues or the collective mode is described by a Gaussian distribution. Analytical results for the ensemble averaged mean and the variance are known \cite{vp2010} in terms of the spectra of $\xi$.

\begin{figure}
        \centering
               \includegraphics [width=0.5\textwidth]{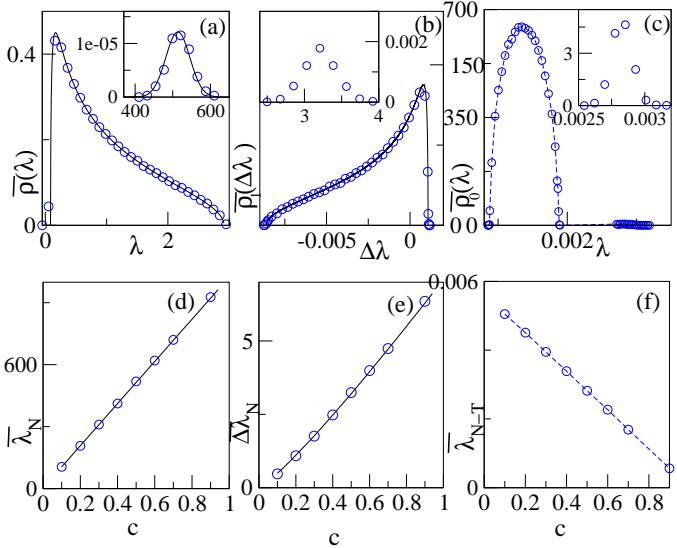}
                \caption{ {The spectral density of the originally non-zero bulk spectra and the well separated emerging spectra of $\mathsf{C}^{(\alpha)}$. In this figure we show results for $N=1024$, $T=512$. We consider the simplest model for a non-random correlation matrix with elements $\xi_{jk}=\delta_{jk}+c\,(1-\delta_{jk})$. In the top figures, viz \ref{deneqc}(a), (b) and (c) we show the bulk densities $\overline{\rho}(\lambda)$, $\overline{\rho}_{1}(\Delta\lambda)$ and $\overline{\rho}_{0}(\lambda)$, respectively for non-zero spectra, non-zero spectral corrections and for emerging spectra, where $c=0.5$. In insets of these figures, numerical densities of the corresponding separated individual eigenvalues are shown. In Fig. \ref{deneqc}(a) we plot the CWOE theory (\ref{eqdeneqc}). In the inset of the same figure we plot the theoretical density using results for the mean and the variance given in \cite{vp2010}. In Fig. \ref{deneqc}(b) we use the ansatz (\ref{denansatz}), by estimating scaling and shifting parameters from (\ref{m1eqc}, \ref{steqc}), with a redefined variance $(1-c)$, to plot the theory. Distribution of separated individual eigenvalues in insets of Figs. \ref{deneqc}(b) and \ref{deneqc}(c) both are not Gaussian. Averaged mean positions as a function of the correlation coefficient $c$ are shown in the bottom figures, Figs. \ref{deneqc}(d), \ref{deneqc}(e) and \ref{deneqc}(f), respectively for the separated individual eigenvalues in $\overline{\rho}(\lambda)$, $\overline{\rho}_{1}(\Delta\lambda)$ and $\overline{\rho}_{0}(\lambda)$. In this figure we use solid lines for the theory and open circles for the numerical results.}}
\label{deneqc}
\end{figure}

\begin{figure}
        \centering
               \includegraphics [width=0.5\textwidth]{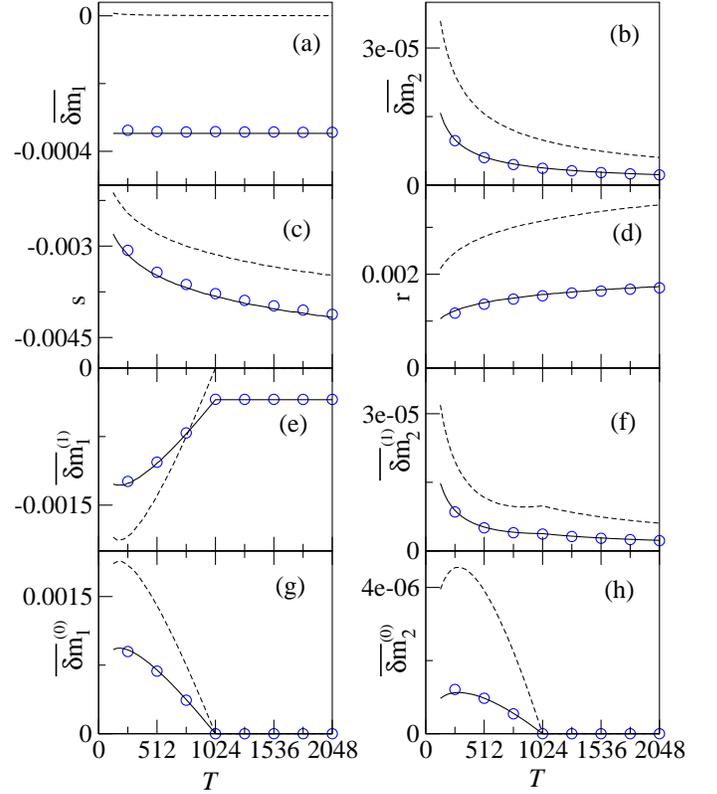}
                \caption{ Comparison of {theoretical and numerical results for CWOE where the nonrandom matrix elements are $\xi_{jk}=\delta_{jk}+c\,(1-\delta_{jk})$. We choose $c=0.5$ and varied $T$ for fixed $N=1024$. This figure repeats the pattern of Fig. \ref{m1m2}. In this figure we compare the moments only for the bulk densities. The null hypothesis, i.e., the corresponding WOE theory is shown by dashed lines.}} 
\label{m1m2eqc}
\end{figure}
To obtain some of the analytical results for the power mapped singular correlation matrices of this model we simply use the fact that for the bulk only the variance is rescaled. This allows us to derive results for the moments of bulk densities. Using $\sigma^{2}=(1-c)$ in Eqs. (\ref{momcjke},\ref{momcjko}) and repeating the calculations of the previous section step by step, we find
\begin{eqnarray}\label{m1eqc}
\overline{\delta m}_{1}&\sim&\alpha(1-c)\,\text{ln}(1-c),
\nonumber\\
\overline{\delta m}_{2}&\sim&\overline{\delta m}_{1}^{2} +
\dfrac{\alpha^{2}(1-c)^{2}}{4\kappa}
\bigg(
[\text{ln}(T)+c_{1}-2\text{ln}(1-c)]^{2}
\nonumber\\
&+&
c_{2}
\bigg).
\end{eqnarray}
Next, extending the ansatz (\ref{denansatz}) for the bulk density $\overline{\rho'}(\lambda)$, we calculate 
\begin{equation}
\overline{\delta m^{(1)}_{1}}=\begin{cases}
s\overline{m}_{1}+r, & \text{for $\kappa\ge1$},
\\
s\overline{m}_{1}+\kappa \,r, & \text{for $\kappa\le1$}.
\end{cases}
\end{equation}
\begin{equation}
\overline{\delta m^{(1)}_{2}}=
\begin{cases}
\overline{m}_{2}s^{2} +r^{2}+2sr\overline{m_{1}},& \text{for $\kappa\ge1$},
\\
\overline{m}_{2}s^{2} +r^{2}\kappa+2sr\overline{m}_{1} & \text{for $\kappa\le1$},
\end{cases}
\end{equation}
where $\overline{m}_{1}$ and $\overline{m}_{2}$ are the first and the second moments $\overline{\rho'}(\lambda)$ which in this case are $(1-c)$ and $(1-c)^{2}(1+1/\kappa)$ respectively and we have replaced the superscript $\{\text{mp}\}$ by $(1)$. For $\kappa\ge1$ and for a given $c$, $s$ and $t$ can be estimated in terms of $\overline{\delta m}_{1}$ and $\overline{\delta m}_{2}$:
\begin{eqnarray}
\label{steqc}
s&=&-\sqrt{\dfrac{\kappa[\overline{\delta m}_{2}-\overline{\delta m}_{1}^{2}]}{(1-c)^{2}}},
\nonumber\\
r&=&\overline{\delta m}_{1}-s(1-c).
\end{eqnarray}
Extrapolating $s$ and $r$ for $\kappa\le1$ we obtain
\begin{equation}
\overline{\delta m^{(1)}_{1}}=\kappa \overline{\delta m_{1}}+(1-c)s(1-\kappa),
\end{equation}
while the result for the second moment is the same as obtained in Eq. (\ref{delm22}).

\begin{figure}[!t]
        \centering
               \includegraphics [width=0.5\textwidth]{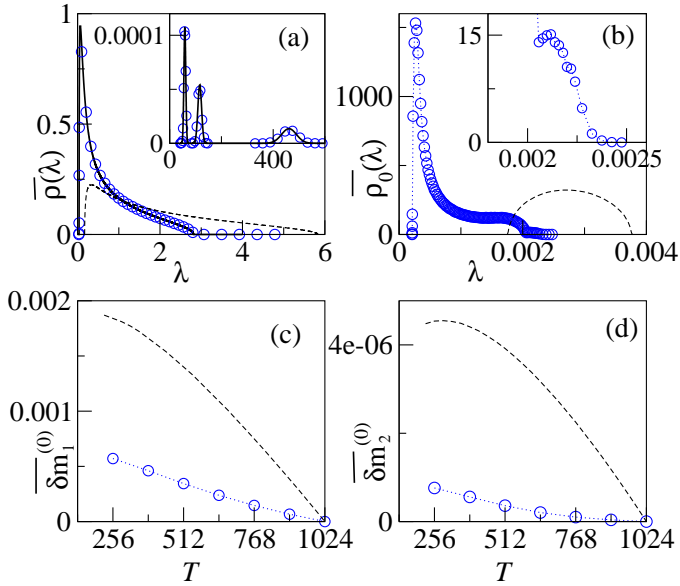}
                \caption{{The spectral density of the originally non-zero bulk spectra and the well separated emerging spectra are shown respectively in Figs. \ref{3blc}(a) and \ref{3blc}(b). The power mapped correlation matrices $\mathsf{C}^{(q)}$ used, result from a CWOE where the averaged correlation matrix $\xi$ has $3$-diagonal blocks. The parameters are chosen as $N=N_{1}+N_{2}+N_{3}=1024$, where $N_{1}=N/2$, $N_{2}=N_{3}=N/4$ and the corresponding correlation coefficients are, respectively, $c_{1}=0.9$, $c_{2}=0.45$ and $c_{3}=0.225$. In the top figures we consider $T=512$ while it is varied in the bottom figures. In the inset of Fig. \ref{3blc}(a) the density of the separated individual eigenvalues are shown in good agreement with the theory. The first two moments, $\overline{\delta m}^{(0)}_{1},\,\overline{\delta m}^{(0)}_{2}$ are shown as function of $T$, respectively, in Fig. \ref{3blc}(c) and in Fig. \ref{3blc}(d). We use solid lines to represent theoretical curves and open circles for the numerical results. We use black dashed lines to represent WOE results.}} 
\label{3blc}
\end{figure}

In Figs. \ref{deneqc}(a), \ref{deneqc}(b) and \ref{deneqc}(c), we show bulk densities for the non-zero eigenvalues, non-zero eigenvalue corrections and emerging spectra of $\mathsf{C}^{(q)}$, respectively, for $c=0.5$. Theory of Fig. \ref{deneqc}(a) comes from Eq. (\ref{eqdeneqc}) while in Fig. \ref{deneqc}(b) it comes from the ansatz for the density of the non-zero eigenvalue corrections, $\overline{\rho}_{1}(\Delta\lambda)$, where $s$ and $r$ are extrapolated for $\kappa\le1$. Theory of Fig. \ref{deneqc}(c) is not known. Densities observed in these figures are different from those for the WOE. In the inset of these figures we show the density of separated individual eigenvalues. {Interestingly, the emerging spectra also show the separation of eigenvalues which is evident from Fig. \ref{deneqc}(c)}. In Figs. \ref{deneqc}(d), \ref{deneqc}(e) and \ref{deneqc}(f), we show the averaged mean of the largest eigenvalues {corresponding} to $\overline{\rho}(\lambda)$, $\overline{\rho}_{1}(\Delta\lambda)$ and $\overline{\rho}_{0}(\lambda)$. If the the separation is large then {we may estimate the averaged mean of the largest eigenvalue corrections by $\overline{\Delta\lambda_{N}}\sim \alpha\overline{\lambda_{N}}\text{log}(\overline{\lambda_{N}^{2}})/2$.} In Fig. \ref{deneqc}(e) we have compared this theory with numerics. {Note here that separated individual eigenvalues appear still on the right side of the bulk.}

\begin{figure}[!t]
        \centering
               \includegraphics [width=0.5\textwidth]{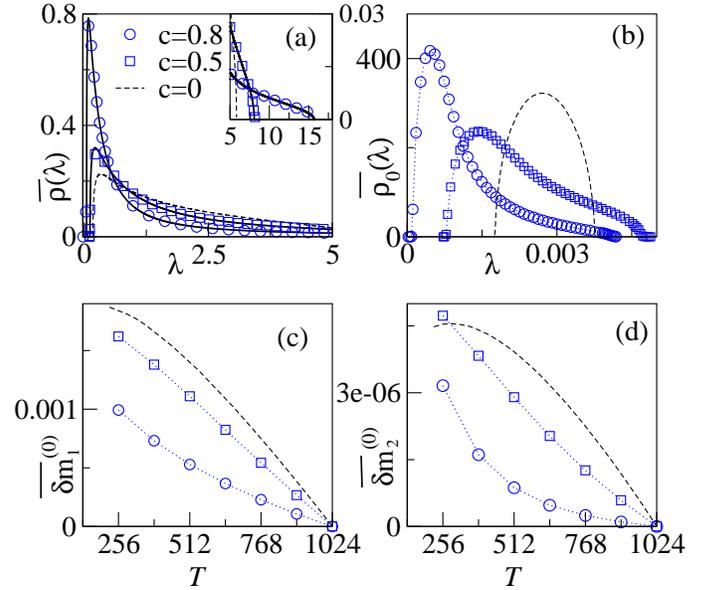}
                \caption{{The spectral density of the originally non-zero bulk spectra and the well separated emerging spectra are shown respectively in Figs. \ref{banded}(a) and \ref{banded}(b). The power mapped correlation matrices $\mathsf{C}^{(q)}$ used, result from a CWOE where the averaged correlation matrix $\xi$ has an exponentially decaying band structure $\xi_{jk}=c^{|j-k|}$.  With an outlay similar to Fig. \ref{3blc} we display
spectral densities and the moments of emerging spectra for the CWOE. We consider here $N=1024$, $\kappa=1/2$ and the correlation coefficient $c=0.8$ ( shown by open circles) and $c=0.5$ (shown by open squares). In the inset of Fig. \ref{banded}(a) tails of the densities are amplified; separated individual eigenvalues are not seen. Black dashed lines are used to represent the corresponding WOE results.}}
\label{banded}
\end{figure}

In Fig. \ref{m1m2eqc} we compare the numerical moments with theoretical results. {We also compare the corresponding theoretical results of WOE (shown in the figure by dashed line). Note that the moments of the emerging spectra are below the corresponding WOE curve. We obtain} qualitatively similar results for $\ell=3$ where we have considered $N_{1}=N/2$, $N_{2}=N_{3}=N/4$ with the nonrandom correlation coefficients in the blocks to be $c_{1}=0.9$, $c_{2}=0.45$ and $c_{3}=0.225$. In Fig. \ref{3blc}(b) we notice a hump in the tail, indicating that the delta functions are about to merge with the bulk.

\subsection{Ensemble correlation matrix with smooth band structure}

{For the emerging spectra in our second example, i.e. $\xi_{jk}=c^{|j-k|}$, we present numerical results while we compare the density of the original non-zero eigenvalues with the theory obtained by solving Eq. (\ref{denCWE}).} In Fig. \ref{banded} we present the results where $N=1024$ and $\kappa=1/2$. For different values of $c=0.8,\,0.5$ and $c=0$. Bulk eigenvalue densities are compared in Fig. \ref{banded} (a) where we find good agreement of the theory with numerics. As shown in this figure, the peak of this distribution located near zero increase with increasing correlations, while the decay of the tail for the density at large values is faster. In Fig. \ref{banded} (b) we compare densities of emerging spectra which changes with $c$. Movements of the upper and the lower spectral edges with $c$ are remarkably mobile in this example, and indeed for larger values of $N/T$ readily the lower edges can take negative values (not shown). In Figs. \ref{banded} (c) and \ref{banded} (d) the first two moments of $\overline{\rho}_{0}(\lambda)$ are compared for different time horizons. These moments qualitatively explain what we see in Fig. \ref{banded} (b). Interestingly, for strong correlations, these moments develop a convex profile which may well be worth further investigations.

\section{Conclusions and outlook}

 Singular correlation matrices arise when the number of time series to be analyzed exceeds their length. This situation is not uncommon when dealing with non-stationary systems, where the correlation matrix has to be estimated for short time horizons. Eigenvalues of correlation matrices are frequently used in various fields \cite{Finance1, Finance2,Finance3,Finance4,diverseAT, gene, Sesmic,Seba:03}. Using more time series will not produce more non-zero eigenvalues if the correlation matrix becomes singular. The central question treated in this paper is if and how more information can be extracted from the additional time series, without access to longer ones. To get access to the subspace corresponding to the zero-eigenvalues, we use the power map and study the spectrum emerging from these zero eigenvalues due to the non-linearity of the map. Note that we not only have found spectral characteristics distinctive of correlations in the emerging spectra, but also some characteristic properties of the eigenvectors appear, as we saw in the case of block-diagonal correlations. Separated individual eigenvalues appear those characterize the blocks, as can be verified also from the eigenvectors pertaining to these eigenvalues. In Sec. II we showed that the power map is also an efficient tool to suppress noise in a correlation analysis involving short time series leading to singular matrices. Thus the same tool can be used to fulfill a standard requirement of a correlation analysis, namely noise suppression and to access the zero-eigenvalue subspace.

We use random matrix ensembles introduced by Wishart to obtain the emerging spectra for random white noise time series that serve as reference for lack of information i.e. as null hypothesis. These ensembles have been generalized to include actual correlations, as correlations constitute the information we seek. We show by means of linear response calculations and numerical analysis that the emerging spectra for such ensembles are very sensitive to the presence or absence of correlations. This is an essential point, as it allows to detect changes in the underlying dynamical structures of a non-stationary system.

Interesting applications for a detection method based on emerging spectra include financial markets, chemical reactors with a high number of probes, observational data from earthquakes resulting from the same fault system or certain biological or medical data like EEG signals. On a more formal note we can hope the problems related to critical statistics \cite{verbaarschot,others} can also be attacked in this way, and indeed specialists in the field have expressed interest to use the power map on chiral ensembles and to regularize Dirac operators \cite{privcom}.

\section{Acknowledgment}
We are thankful to F. Leyvraz, J. J. M. Verbaarschot and T. Guhr for important and useful discussions. We acknowledge financial support from the project 44020 by CONACyT, Mexico, and project PAPIIT UNAM RR 113311, Mexico. Vinayak is a postdoctoral fellow in DGAPA/UNAM.

\appendix

\section{Details of the portfolio optimization study} \label{app:opt}

The weight vector of the minimal variance portfolio is calculated as
\begin{equation}
w_\mathrm{opt}=\frac{\Sigma^{-1} e}{e^t \Sigma^{-1} e} \ ,
\label{eq:wopt}
\end{equation}
where $e$ is a vector of length $N$ with all entries set to one, $e^t$ denotes the transposed vector. 
In order to calculate the optimal weights (\ref{eq:wopt}), we need to know the covariance matrix $\Sigma$ of the $N$ stock returns.
In practice, this covariance matrix has to be estimated using historical data.
Here we consider a model setting with $N=100$ stocks, a model correlation matrix $C_0$ with 5 blocks of size 20, corresponding to industry sectors, and randomly distributed but fixed standard deviations $\sigma_k$.
In a factor model that reflects our correlation matrix $C_0$, we simulate time series of length $N$, estimate the correlation matrix $C^\mathrm{(samp)}$ and apply the power map to arrive at the matrix $C^{(q)}$ with entries
\begin{equation}
C^{(q)}_{kl}={\rm sign}\,C^\mathrm{(samp)}_{kl} \, \big| C^\mathrm{(samp)}_{kl} \big|^q \ .
\end{equation}
We multiply with the standard deviations to get the elements of the covariance matrix,
\begin{equation}
\widehat{\Sigma}_{kl}=\sigma_k \sigma_l C^{(q)}_{kl} \ .
\end{equation}
With this covariance matrix we calculate the weights 
\begin{equation}
\widehat{w}_\mathrm{opt}=\frac{\widehat{\Sigma}^{-1} e}{e^t \widehat{\Sigma}^{-1} e} \ .
\label{eq:wopthat}
\end{equation}
Using the model correlation matrix $C_0$ and the corresponding covariance matrix $\Sigma_0=\sigma C_0 \sigma$, with $\sigma={\rm diag}(\sigma_1, \ldots, \sigma_N)$, the actual portfolio variance for the weights (\ref{eq:wopthat}) reads
\begin{equation}
\Omega^2=\widehat{w}_\mathrm{opt}^{\,t} \Sigma_0 \widehat{w}_\mathrm{opt} \ ,
\end{equation}
whereas the minimal variance $\Omega_0^2$ is obtained by calculating the optimal weights (\ref{eq:wopt}) for $\Sigma_0$.


\begin{thebibliography}{99}
\bibitem{comment} The Wishart ensemble is the obvious choice for an ensemble of zero cross-correlation matrices, because in the infinite time limit all elements of the ensemble have zero cross-correlations. For finite time we can determine the probability for any given correlation with some narrow distribution and thus find the likelihood as compared to the Wishart ensemble as "null hypothesis". In practice we do not need to do this in the present paper, as we also use ensembles of correlated matrices and therefore can compare distributions which, at least numerically, we can get to arbitrary exactitude.
\end{thebibliography}
\end{document}